\documentclass[10pt]{article}

\usepackage[letterpaper,margin=0.75in,columnsep=0.25in]{geometry}
\usepackage[none]{hyphenat}
\usepackage{newtxmath}
\usepackage{amsfonts}
\usepackage{enumitem}
\usepackage{tabularx}
\usepackage{algorithm}
\usepackage{algorithmic}
\usepackage{booktabs}
\usepackage{xspace}
\usepackage[table]{xcolor}
\usepackage{multirow}
\usepackage{tcolorbox}
\usepackage[numbers,square,sort&compress]{natbib}
\setcitestyle{square,numbers,comma}
 \usepackage{amsmath}
\usepackage{xurl}
\usepackage[hidelinks,breaklinks=true]{hyperref}
\usepackage{microtype}
\newcommand{\srf}{\textsc{SRF}}
\newcommand{\pia}{\textsc{PIA}}
\newcommand{\dhi}{\textsc{DHI}}
\newcommand{\bci}{\textsc{BCI}}
\newcommand{\rcf}{\textsc{RCF}}
\newcommand{\cfc}{\textsc{CFC}}
\newcommand{\ccv}{\textsc{CCV}}
\newcommand{\ssr}{\textsc{SSR}}

\newcommand{\keywords}[1]{\par\smallskip\noindent\textbf{Keywords:} #1\par\smallskip}

\title{The Patchwork Problem in LLM-Generated Code}

\author{
\begin{tabular}{c}
Viraaji Mothukuri, Reza M. Parizi\\
Decentralized Science Lab, College of Computing and Software Engineering\\
Kennesaw State University, GA, USA\\
\texttt{vmothuku@students.kennesaw.edu} \quad \texttt{rparizi1@kennesaw.edu}
\end{tabular}
}

\date{}

\begin{document}

\twocolumn[
\maketitle

\begin{abstract}
LLM-generated code often compiles, passes tests, and appears correct, yet breaks once deployed. The root cause is frequently structural rather than logical. A generated endpoint references configuration keys never declared in the project, an import targets a package that does not exist in any registry, or a new route omits the authentication guard applied to every sibling endpoint. Each patch is locally valid but globally incoherent, and standard CI toolchains rarely surface these failures. As LLM-powered coding tools see widespread adoption, this blind spot poses a growing risk to software quality. We call this the \textbf{patchwork problem}. This paper formalizes structural coherence as consistency invariants over graph representations of repository artifacts, including import, call, dependency, configuration, schema, resource, control-flow, and routing graphs, and introduces an eight-category failure taxonomy distinguishing defects specific to LLM generation from those merely amplified by it. We present a hybrid verification framework that delegates to mature static analysis tools where they already excel and deploys purpose-built detectors for cross-cutting invariants underserved by existing toolchains, targeting provable constraint violations rather than heuristic pattern matching. Empirical evaluation across two frontier models under four prompting strategies reveals that the vast majority of structural failures evade type checking, testing, and SAST entirely, and that failure patterns diverge qualitatively between models in ways that challenge model-agnostic mitigation strategies. External validation on real-world AI-generated repositories confirms that these failures are not artifacts of controlled experimentation but are prevalent wherever LLMs write code with minimal human oversight.
\end{abstract}

\keywords{LLM code generation, structural coherence, static analysis, graph invariants, code quality, neural code synthesis.}

\vspace{1em}
]

\section{Introduction} \label{sec:intro}
Code generation from Large Language Models has achieved remarkable results on isolated programming tasks \cite{jiang2024survey,chen2021codex,li2022alphacode, austin2021program}, driving rapid adoption, with millions of engineers using LLM-powered assistants daily. Yet a gap persists between benchmark performance and production utility \cite{jimenez2023swebench}. Code that appears correct in isolation frequently fails when integrated into real software systems \cite{zhang2024llmhallucinations}, and the dominant failure mode is structural rather than functional. A generated patch may compile, pass type checking, and satisfy local tests while violating invariants that span the repository. Consider a FastAPI endpoint referencing a Pydantic model with hallucinated field names, or a Django view assuming environment variables that are never declared in the project configuration. Such patches exhibit local correctness but global incoherence: they pass the checks developers rely on and fail only when exercised in the context of the full system.

Current evaluation methodologies do not surface these failures systematically. Type checking and linting often miss semantic inconsistencies that cross file boundaries. Test suites cannot cover every integration point. SAST tools typically focus on taint flows rather than structural coherence. The result is a blind spot in which generated code enters codebases carrying latent defects that remain invisible to standard toolchains. We term this the \textbf{patchwork problem}. LLM-generated patches may be individually well-formed yet fail to cohere into a consistent whole, particularly at repository scale, where consistency constraints span imports, dependencies, configurations, schemas, and security contracts \cite{liu2024repobench}.

Our approach formalizes structural coherence as consistency invariants over graph representations of repository artifacts \cite{yamaguchi2014cpg}. A key design insight is that reliable detection requires matching each invariant class to an appropriate verification strategy. Categories where mature static analysis tools already capture the relevant language semantics can be delegated to those tools, while categories that require cross-graph reasoning, absent from existing toolchains, call for purpose-built detectors that target constraint violations under explicit assumptions and produce actionable evidence. This work makes three contributions: 
(1) We introduce a \textbf{taxonomy of eight structural failure categories}, each defined by graph-based consistency invariants, distinguishing failures characteristic of LLM-generated patches from issues merely amplified by them. (2) We present a \textbf{multi-graph verification framework} that integrates mature static analysis tools (\texttt{mypy}, \texttt{tsc}, \texttt{pylint}, \texttt{ESLint}) with purpose-built cross-graph detectors over eight repository graphs, producing localized evidence traces for each violation.
(3) We provide an \textbf{empirical study across 336 generations} from two frontier models under four prompting conditions, along with external validation on 43 real-world AI-generated repositories.

\section{Related Work}
\label{sec:related}

Table \ref{tab:positioning} situates our contribution relative to prior work across four research threads. Hallucination characterization work \cite{zhang2024llmhallucinations, spracklen2024packagehallucinations} establishes the empirical prevalence of structural defects in generated code but characterizes them descriptively rather than as verifiable constraint violations. Repository-level benchmarks, including RepoBench \cite{liu2024repobench}, SWE-bench \cite{jimenez2023swebench}, EvoCodeBench \cite{li2024evocodebench}, BaxBench \cite{vero2025baxbench}, SecRepoBench \cite{dilgren2025secrepobench}, SecureVibeBench \cite{chen2025secureagentbench}, and SWE-agent \cite{yang2024sweagent} demonstrate that snippet-level performance does not transfer reliably to repository-scale tasks, yet their evaluation criteria remain outcome-oriented (test passage, exploit success) rather than diagnosing which structural invariants are violated. Graph-based representations such as Code Property Graphs \cite{yamaguchi2014cpg}, CODE-MVP \cite{wang2022codemvp}, and GALLa \cite{zhang2025galla} leverage graphs as representational substrates but do not operationalize them as a constraint verification layer. Secure generation approaches, including CodeGuard+ \cite{fu2024codeguardplus} and SafeGenBench \cite{li2025safegenbench}, target vulnerability prevention without formalizing structural coherence. Our work addresses this gap by defining structural incoherence as violated consistency constraints across graph representations and producing localized evidence traces that attribute failures to specific constraint violations.

\begin{table}[!ht]
\centering
\tiny
\caption{Summary of related work}
\begin{tabularx}{\linewidth}{p{1cm}|p{3.2cm}|p{3.2cm}}
\toprule
\textbf{Work} & \textbf{Focus} & \textbf{Gap} \\
\midrule
\cite{zhang2024llmhallucinations} & Taxonomy of hallucinations in repo-level generation & Descriptive; no formal invariants or automated detection \\
\midrule
\cite{spracklen2024packagehallucinations} & Quantitative measurement of fabricated package references & Dependency-only; no cross-artifact verification \\
\hline
\cite{liu2024repobench} & Cross-file code completion with retrieval subtasks & Evaluates completion, not structural coherence \\
\midrule
\cite{jimenez2023swebench} & GitHub issue resolution via test passage & Test-based; misses failures that evade tests \\
\midrule
\cite{li2024evocodebench} & Repo-aligned generation with dependency annotations & Pass@k metric; no structural invariant checking \\
\midrule
\cite{vero2025baxbench} & Backend generation with exploit-based security assessment & Exploit-focused; no config or schema verification \\
\midrule
\cite{dilgren2025secrepobench} & Security-focused patch evaluation & Vulnerability labels, not structural constraints \\
\midrule
\cite{chen2025secureagentbench} & Multi-file vulnerability scenarios & Security-scoped; no general structural coverage \\
\hline
\cite{yang2024sweagent} & Autonomous multi-step repo editing & Evaluates resolution rate, not edit coherence \\
\hline
\cite{yamaguchi2014cpg} & Unified code property graph for vulnerability discovery & Single-file; no cross-file or config constraints \\
\midrule
\cite{wang2022codemvp} & Multi-view contrastive pre-training & Graphs as training signal, not verification layer \\
\midrule
\cite{zhang2025galla} & Graph-aligned fine-tuning for structural semantics & Alignment target, not constraint checking \\
\hline
\cite{fu2024codeguardplus} & Constrained decoding for secure generation & Decoding-time; no post-hoc structural verification \\
\midrule
\cite{li2025safegenbench} & Dual-judge vulnerability detection & Vulnerability-scoped; no structural coherence \\
\bottomrule
\end{tabularx}
\label{tab:positioning}
\end{table}

\section{Structural Failure Taxonomy} \label{sec:taxonomy}
The patchwork problem manifests through \textbf{structural failures}, defined as violations of consistency invariants at the repository scale, verifiable via static graph analysis without execution. These failures differ from functional bugs in that individual patches appear correct yet collectively violate project contracts. Our taxonomy comprises eight categories, each defined by formal invariants, required graph artifacts, and failure characteristics specific to LLM outputs. Table~\ref{tab:taxonomy_summary} summarizes the classification.

\textbf{Symbol Resolution Failures (\srf{})}: A symbol resolution failure occurs when a referenced name cannot be resolved within the repository's module graph. Formally, for symbol reference $r$ in file $f$ with module graph $\mathcal{M}$, the invariant requires $\forall r \in \text{refs}(f): \exists d \text{ s.t. } \text{resolve}(r, f, \mathcal{M}) \rightarrow d$. Detection requires the import graph and symbol table, optionally augmented with type information for generic resolution. Evidence traces record file, line, symbol, expected module, and resolution outcome. This failure class is \textit{amplified} in LLM outputs because models operating with incomplete context frequently invent plausible but non-existent module names or reference deprecated APIs.

\textbf{Phantom Internal API (\pia{})}: Phantom API failures occur when generated code invokes internal functions with incorrect signatures or semantics inconsistent with declared interfaces. The invariant requires signature compatibility such that for call site $c$ invoking symbol $s$ with signature registry $\Sigma$, $\text{compatible}(\text{sig}(c), \Sigma(s)) = \text{true}$. Detection leverages the call graph and signature registry extracted from type annotations and protocol declarations. This category is \textit{strongly amplified} as LLMs hallucinate method signatures based on naming conventions rather than actual declarations, particularly for internal APIs underrepresented in training data.

\textbf{Dependency Hallucination (\dhi{})}: Algorithm~\ref{alg:dhi} validates that all external imports reference packages declared in the project's dependency manifests. External imports not found in the dependency graph trigger registry queries to PyPI or npm, distinguishing fully hallucinated packages (nonexistent in registries) from undeclared-but-existing packages. For npm packages, import names match registry names directly. For PyPI, the detector assumes direct correspondence between import names and package names, which holds for the majority of packages but not for cases where import and distribution names diverge (e.g., \texttt{yaml} vs. \texttt{PyYAML}, \texttt{cv2} vs. \texttt{opencv-python}). The resulting phantom module set feeds downstream into \srf{} and \pia{} detection. 

\textbf{Build/Configuration Incoherence (\bci{})}: Build configuration failures arise when generated code assumes configurations inconsistent with the repository's declared state. The invariant requires that for each configuration assumption $a$ implied by generated code, a satisfying declaration exists in the configuration space $\mathcal{C}$. Detection operates over the build graph and configuration graph encompassing entrypoints, environment variables, and framework settings. Four invariant classes apply across languages, namely entrypoint existence, environment variable declaration, module system consistency, and framework configuration alignment. This category is \textit{amplified} under weak retrieval context where models default to standard configurations divergent from project settings.

\textbf{Resource Coherence Failures (\rcf{})}: Resource coherence failures occur when code fails to provide declared resources, encompassing both filesystem resources and computational contracts. Filesystem resource failures arise when code references files, assets, templates, or migrations that do not exist, with the existence invariant requiring $\text{exists}(\text{resolve\_path}(r, \text{root})) = \text{true}$ for each resource reference $r$ and additional ordering constraints for sequential resources such as database migrations. Return contract failures arise when functions with declared return types contain execution paths that do not produce a value matching the declared type, violating the contract that the function's signature promises to callers. Schema completeness failures arise when model definitions omit fields required by consuming code. Detection operates over the resource graph mapping code references to filesystem paths, the CFG for return-path reachability analysis, and the schema graph for field completeness validation. LLMs exhibit \textit{amplified} failure rates across all three sub-categories by generating plausible paths based on conventions rather than actual repository structure, omitting return statements on error-handling branches, and producing incomplete schema definitions.

\textbf{Control Flow Coherence (\cfc{})}: Control flow failures manifest as CFG anomalies including unreachable blocks, contradictory conditions, exception flow misuse, and dead error handling. Invariants require full reachability from entry ($\forall v: \text{reachable}(v_0, v)$) and exception handler type consistency. Detection operates on intraprocedural CFGs with exception edge annotations. While unreachable code is a general bug class, LLMs exhibit \textit{amplified characteristic patterns} such as overbroad exception handling, redundant null checks, and copy-paste control flow inconsistencies.

\textbf{Cross-File Contract Violations (\ccv{})}: Contract violations occur when producer and consumer modules exhibit interface mismatches across file boundaries, including wrong field names, incompatible serialization, incorrect error codes, and misaligned assumptions. The invariant requires schema compatibility whereby $\text{schema}(P.\text{output}) \supseteq \text{schema}(C.\text{input})$ with type consistency. Detection requires the call graph with edges spanning files and the schema graph extracted from OpenAPI specs, Pydantic models, or Zod schemas. LLMs \textit{amplify} this failure by hallucinating response fields based on naming conventions.

\begin{figure*} [ht]
    \centering
    \includegraphics[width=\linewidth]{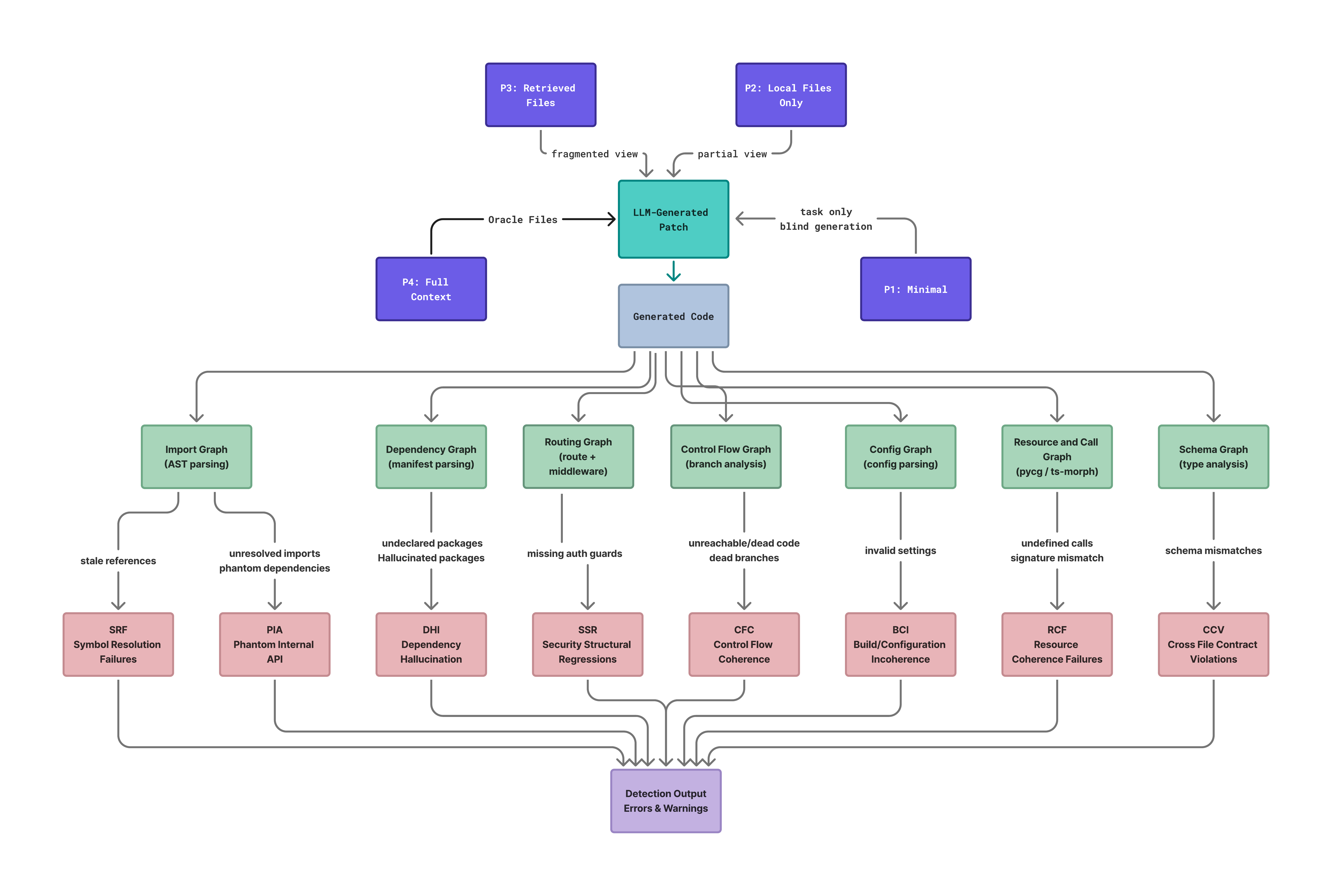}
    \caption{Overview of the Proposed Framework}
    \label{fig:patch}
\end{figure*}

\textbf{Security Structural Regressions (\ssr{})}: Security structural regressions occur when application wiring violates security contracts absent classic taint flow vulnerabilities. The invariant requires guard coverage such that $\forall r \in R: \text{guarded\_by}(r, M)$ for security-critical routes $R$ and required guards $M$. Detection requires routing and middleware attachment graphs specific to each framework. We scope detection to FastAPI (dependency injection guards), Django (permission decorators), Express (middleware chains), and Next.js (middleware matchers). LLMs \textit{amplify} this failure by wiring middleware incorrectly, even when producing syntactically correct code.

\begin{table*}[t]
\centering
\tiny
\caption{Structural failure taxonomy}
\begin{tabular}{@{}llllcc@{}}
\toprule
\textbf{ID} & \textbf{Category} & \textbf{Primary Graph(s)} & \textbf{LLM Profile} & \textbf{Type check evasion} & \textbf{Test Evasion} \\
\midrule
\srf{} & Symbol Resolution & Import + Symbol Table & Amplified & Partial & Partial \\
\pia{} & Phantom Internal API & Call + Signature & Strongly Amplified & Yes & Yes \\
\dhi{} & Dependency Hallucination & Dependency + Registry & Specific to LLMs & Yes & Yes \\
\bci{} & Build/Config Incoherence & Build + Config & Amplified & Yes & Yes \\
\rcf{} & Resource Coherence & Resource + CFG + Schema & Amplified & Yes & Yes \\
\cfc{} & Control Flow Coherence & CFG & General/Amplified & Partial & Partial \\
\ccv{} & Cross-File Contracts & Call + Schema & Amplified & Yes & Yes \\
\ssr{} & Security Structural & Routing + Middleware & Amplified & Yes & Yes \\
\bottomrule
\end{tabular}

\label{tab:taxonomy_summary}
\end{table*}

\section{Verification Framework}
\label{sec:framework}

\textbf{Architecture Overview}: The verification framework employs a hybrid architecture guided by a precision-first design philosophy in which each taxonomy category is matched to the verification strategy that maximizes detection precision for its invariant class. Categories where mature static analysis tools already handle the relevant language semantics (symbol resolution, signature compatibility) delegate to those tools, inheriting their years of edge-case handling. Control flow coherence employs a hybrid approach combining custom graph reachability and pattern matching with SAST tool delegation, reflecting the observation that no single layer achieves adequate coverage alone. Categories requiring cross-graph reasoning absent from existing toolchains (configuration incoherence, dependency hallucination, security regressions, resource coherence, cross-file contracts) employ purpose-built detectors that target provable constraint violations rather than heuristic pattern matching. This division reflects the empirical observation that reimplementing established analyses from scratch produces low-precision detectors due to the long tail of language-specific edge cases (exception flows, generators, context managers, async patterns), while the novel cross-cutting invariants central to the patchwork problem have no existing tool coverage. The framework operates on each repository state after generation, constructing graph representations from the combined original and generated code and routing each to the appropriate verification backend. The framework's output for each finding is a localized evidence trace recording the violated invariant, the implicated files and line numbers, and the constraint that would need to hold for the code to be structurally sound. All source code, graph construction scripts, detection pipelines, evaluation configurations, and external validation datasets required to reproduce our results are publicly available \cite{patchwork-repo}. Figure~\ref{fig:patch} illustrates the end-to-end verification pipeline, showing how graph construction connects prompting conditions to failure detection.

\textbf{Graph Construction}: For each repository state after generation, the framework constructs eight graph representations spanning structural, behavioral, and configurational dimensions. Table~\ref{tab:graphs} summarizes the construction method for each graph type.

\begin{table}[h]
\centering
\tiny
\caption{Graph construction methods by language}
\begin{tabular}{@{}p{1.4cm}p{2.8cm}p{2.8cm}@{}}
\toprule
\textbf{Graph} & \textbf{Python} & \textbf{TypeScript} \\
\midrule
Import & \texttt{ast} module; relative import resolution & \texttt{ts.createProgram} with \texttt{tsconfig.json} resolution \\
Call & \texttt{pycg} flow-insensitive points-to analysis & Compiler API type-directed resolution \\
Dependency & \texttt{pyproject.toml}, \texttt{requirements.txt}, \texttt{poetry.lock} + PyPI validation & \texttt{package.json}, \texttt{package-lock.json}, \texttt{yarn.lock} + npm validation \\
Schema & Pydantic \texttt{BaseModel}, SQLAlchemy \texttt{Column} & Zod \texttt{z.object()}, Prisma \texttt{schema.prisma} \\
Config & \texttt{.env}, \texttt{.env.example}, Docker Compose, framework settings & Same sources \\
Resource & \texttt{open()}, \texttt{pathlib.Path}, template loaders, migration deps & Equivalent TS patterns \\
CFG & \texttt{ast} branch analysis; \texttt{pylint} delegation & \texttt{ts-morph} branch analysis; \texttt{ESLint} delegation \\
Routing & FastAPI route decorators, Django URL conf & Express router chains, Next.js middleware matchers \\
\bottomrule
\end{tabular}
\
\label{tab:graphs}
\end{table}

\begin{figure}[h]
\centering
\includegraphics[width=0.7\linewidth]{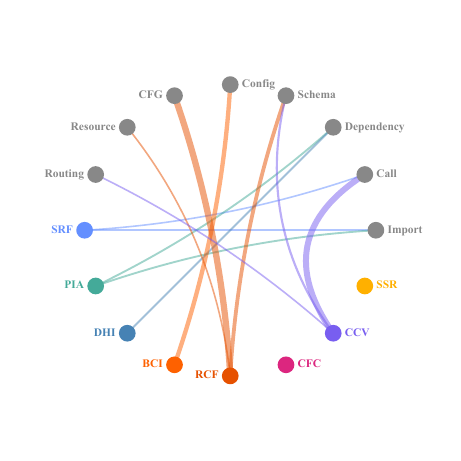}
\caption{Graph-to-category mapping}
\label{fig:chord}
\end{figure}

Figure~\ref{fig:chord} visualizes the many-to-many mapping between graph representations and failure categories, illustrating why the framework requires multiple coordinated analyses rather than a single monolithic pass. Chords connect each graph representation (left) to the failure categories it enables detecting (right). Some categories require multiple graphs, and some graphs serve multiple categories, motivating the hybrid architecture.

\textbf{Detection Algorithms}: The following paragraphs formalize detection for each taxonomy category. Purpose-built detectors (Algorithms~\ref{alg:bci}--\ref{alg:srfpia}, \ref{alg:rcf}, \ref{alg:ccv}, and~\ref{alg:ssr}) target provable constraint violations, while control flow coherence (Algorithm~\ref{alg:cfc}) combines custom analysis with SAST tool delegation. Detection order reflects data dependencies, with \dhi{} running first to produce the phantom module set consumed by \srf{} and \pia{}.

\textbf{Configuration Incoherence Detection (\bci{})}: Algorithm~\ref{alg:bci} detects provable runtime failures from unguarded environment variable accesses. It extracts strict access patterns that throw on missing values (\texttt{os.environ["KEY"]} in Python, unguarded \texttt{process.env.KEY} in TypeScript), eliminates accesses protected by guards (\texttt{try/except}, membership tests, fallback operators), and validates remaining accesses against the repository's configuration space. Arithmetic expressions and safe access patterns with explicit defaults are excluded at extraction time. Every reported finding represents a configuration access that will produce a runtime crash if the variable is absent.

\begin{algorithm}[h]
\caption{Configuration Incoherence Detection (\bci{})}
\small
\label{alg:bci}
\begin{algorithmic}[1]
\REQUIRE Repository $R$ with generated code $G$, config files $C$
\ENSURE Set of validated configuration incoherence findings $F$

\STATE \textbf{Extract unsafe accesses}
\STATE $A \leftarrow \emptyset$
\FOR{each file $f$ in $G$}
    \IF{$f$ is Python}
        \STATE $A \leftarrow A \cup \{\texttt{os.environ["K"]}$ patterns in $f\}$
    \ELSIF{$f$ is TypeScript}
        \STATE $A \leftarrow A \cup \{$unguarded \texttt{process.env.K} in $f\}$
    \ENDIF
\ENDFOR

\STATE \textbf{Guard elimination}
\FOR{each access $a \in A$}
    \IF{$a$ inside \texttt{try/except KeyError} \OR preceded by membership test \OR has \texttt{||}/\texttt{??} fallback}
        \STATE $A \leftarrow A \setminus \{a\}$
    \ENDIF
\ENDFOR

\STATE \textbf{Config-space validation}
\STATE $\mathcal{C} \leftarrow$ keys from \texttt{.env}, \texttt{.env.example}, \texttt{docker-compose.yml}, settings
\STATE $F \leftarrow \emptyset$
\FOR{each access $a \in A$ with key $k$}
    \IF{$k \notin \mathcal{C}$}
        \STATE $F \leftarrow F \cup \{(a, k, \text{file}, \text{line})\}$
    \ENDIF
\ENDFOR
\RETURN $F$
\end{algorithmic}
\end{algorithm}

\textbf{Dependency Hallucination Detection (\dhi{})}: Algorithm~\ref{alg:dhi} validates that all external imports reference packages declared in the project's dependency manifests. External imports not found in the dependency graph trigger registry queries to PyPI or npm, distinguishing fully hallucinated packages (nonexistent in registries) from undeclared-but-existing packages. The resulting phantom module set feeds downstream into \srf{} and \pia{} detection.

\begin{algorithm}[h]
\caption{Dependency Hallucination Detection (\dhi{})}
\small
\label{alg:dhi}
\begin{algorithmic}[1]
\REQUIRE Repository $R$ with manifests $M$, generated code $G$
\ENSURE Set of dependency hallucination findings $F$

\STATE $D \leftarrow$ parse packages from $M$ (\texttt{pyproject.toml}, \texttt{requirements.txt}, \texttt{package.json})

\STATE \textbf{Extract and validate external imports}
\STATE $F \leftarrow \emptyset$
\FOR{each import $i$ in $G$ referencing package $p$}
    \IF{$p \notin$ stdlib \AND $p \notin$ local modules \AND $p \notin D$}
        \STATE Query registry $\mathcal{R}$ (PyPI/npm) for $p$
        \IF{$p \notin \mathcal{R}$}
            \STATE $F \leftarrow F \cup \{(i, p, \text{``hallucinated''})\}$
        \ELSE
            \STATE $F \leftarrow F \cup \{(i, p, \text{``undeclared''})\}$
        \ENDIF
    \ENDIF
\ENDFOR
\RETURN $F$
\end{algorithmic}
\end{algorithm}

\textbf{Symbol Resolution and Phantom API Detection (\srf{}, \pia{})}: Algorithm~\ref{alg:srfpia} leverages the phantom module set from Algorithm~\ref{alg:dhi}. A \texttt{LocalModulePattern} filter excludes intra-generation cross-references and intentional placeholders to prevent false positives from multi-file generation tasks.

\begin{algorithm}[h]
\caption{Symbol Resolution (\srf{}) and Phantom API (\pia{}) Detection}
\small
\label{alg:srfpia}
\begin{algorithmic}[1]
\REQUIRE Import graph $\mathcal{I}$, call graph $\mathcal{K}$, phantom set $P$ from Algorithm~\ref{alg:dhi}
\ENSURE Sets of SRF findings $F_S$ and PIA findings $F_P$

\STATE $F_S \leftarrow \emptyset$, $F_P \leftarrow \emptyset$
\FOR{each import edge $(f, m, s)$ in $\mathcal{I}$}
    \IF{($m \in P$ \OR $\neg\text{resolve}(s, m)$) \AND $m \notin$ \texttt{LocalModulePattern}}
        \STATE $F_S \leftarrow F_S \cup \{(f, m, s)\}$
    \ENDIF
\ENDFOR
\FOR{each call edge $(f, m.s, \text{args})$ in $\mathcal{K}$}
    \IF{$m \in P$ \AND $m \notin$ \texttt{LocalModulePattern}}
        \STATE $F_P \leftarrow F_P \cup \{(f, m, s, \text{args})\}$
    \ENDIF
\ENDFOR
\RETURN $F_S, F_P$
\end{algorithmic}
\end{algorithm}

\textbf{Resource Coherence Detection (\rcf{})}: Algorithm~\ref{alg:rcf} targets three sub-categories. Return contract violations construct intraprocedural CFGs for functions with declared return types and identify execution paths that terminate without producing a value, excluding branches guarded by exception handlers that re-raise or call \texttt{sys.exit}. Filesystem resource violations check that referenced paths (templates, migrations, assets) resolve to existing files. Schema completeness violations flag consuming code that accesses fields absent from the producing model's definition. Every finding represents a provable violation: a reachable path missing a declared return, a path literal that does not resolve, or a field access targeting an undefined name.

\begin{algorithm}[h]
\caption{Resource Coherence Detection (\rcf{})}
\small
\label{alg:rcf}
\begin{algorithmic}[1]
\REQUIRE CFGs $\mathcal{F}$, resource graph $\mathcal{R}$, schema graph $\mathcal{S}$, generated code $G$
\ENSURE Set of resource coherence findings $F$
\STATE $F \leftarrow \emptyset$
\FOR{each function $f$ in $G$ with declared return type $T$}
    \STATE Build intraprocedural CFG; for each path $\pi$ to exit, if $\pi$ has no \texttt{return} of type $T$ and is not exception-terminated, add $(f, \pi, T)$ to $F$
\ENDFOR
\FOR{each resource ref $r$ with path $p$ in $G$}
    \IF{$\neg\text{exists}(\text{resolve}(p, \text{root}))$}
        \STATE $F \leftarrow F \cup \{(r, p)\}$
    \ENDIF
\ENDFOR
\FOR{each consumer access $c.field$ produced by model $M$}
    \IF{$field \notin \text{fields}(M)$}
        \STATE $F \leftarrow F \cup \{(c, M, field)\}$
    \ENDIF
\ENDFOR
\RETURN $F$
\end{algorithmic}
\end{algorithm}

\textbf{Cross-File Contract Violation Detection (\ccv{})}: Algorithm~\ref{alg:ccv} detects interface mismatches across module boundaries via call graph and schema graph analysis. It targets four patterns: disconnected middleware (registered but never imported in route modules), unused decorators referencing nonexistent permission classes, duplicate middleware registrations causing double execution, and field naming mismatches between producer response schemas and consumer access patterns (e.g., \texttt{user\_name} vs. \texttt{username}). Each finding identifies a statically verifiable disconnect between two code locations that must agree for correct execution.

\begin{algorithm}[h]
\caption{Cross-File Contract Violation Detection (\ccv{})}
\small
\label{alg:ccv}
\begin{algorithmic}[1]
\REQUIRE Call graph $\mathcal{K}$, schema graph $\mathcal{S}$, middleware config $\mathcal{M}$, generated code $G$
\ENSURE Set of contract violation findings $F$
\STATE $F \leftarrow \emptyset$
\FOR{each middleware $m$ registered in $\mathcal{M}$}
    \IF{$m \notin$ import edges of any route module}
        \STATE $F \leftarrow F \cup \{(m, \text{disconnected})\}$
    \ENDIF
    \IF{$m$ appears $>$1 time in registration}
        \STATE $F \leftarrow F \cup \{(m, \text{duplicate})\}$
    \ENDIF
\ENDFOR
\FOR{each cross-file call edge $(producer, consumer)$ in $\mathcal{K}$}
    \STATE $S_P \leftarrow$ output fields of $producer$; $S_C \leftarrow$ accessed fields in $consumer$
    \IF{$\exists\, f \in S_C : f \notin S_P$}
        \STATE $F \leftarrow F \cup \{(producer, consumer, S_C \setminus S_P)\}$
    \ENDIF
\ENDFOR
\RETURN $F$
\end{algorithmic}
\end{algorithm}

\textbf{Control Flow Coherence Detection (\cfc{})}: Algorithm~\ref{alg:cfc} employs a hybrid three-layer approach with findings deduplicated by line number. Layer 1 performs BFS reachability from function entry nodes, flagging only entirely dead functions. Layer 2 applies pattern matching for dead code after terminators, tautological conditions, duplicate handlers, and infinite loops. Layer 3 delegates to \texttt{pylint}/\texttt{ESLint} with post-processing filters suppressing known false positives from context managers, generators, and heavy exception scaffolding.

\begin{algorithm}[h]
\caption{Control Flow Coherence Detection (\cfc{})}
\small
\label{alg:cfc}
\begin{algorithmic}[1]
\REQUIRE Generated code files $G$, constructed CFGs $\mathcal{F}$
\ENSURE Set of control flow findings $F$

\STATE $F \leftarrow \emptyset$
\FOR{each file $f$ in $G$}
    \STATE \textbf{Graph-based reachability (Layer 1)}
    \FOR{each function CFG $g \in \mathcal{F}(f)$}
        \STATE $R \leftarrow$ BFS from entry node of $g$
        \IF{all body nodes $\notin R$}
            \STATE $F \leftarrow F \cup \{(f, g, \text{``dead function''})\}$
        \ENDIF
    \ENDFOR
    \STATE \textbf{Pattern-based detection (Layer 2)}
    \STATE $F \leftarrow F \cup$ detect dead-code-after-terminator in $f$
    \STATE $F \leftarrow F \cup$ detect tautological conditions in $f$
    \STATE $F \leftarrow F \cup$ detect duplicate except/switch-case in $f$
    \STATE $F \leftarrow F \cup$ detect infinite loops without exit in $f$
    \STATE \textbf{SAST delegation (Layer 3)}
    \STATE $r \leftarrow$ invoke \texttt{pylint}/\texttt{ESLint} on $f$ (skip if syntax error)
    \STATE Remove findings in context manager, generator, or heavy \texttt{try/finally} contexts
    \STATE $F \leftarrow F \cup r$
    \STATE \textbf{Deduplicate} $F$ by line number (priority: graph $>$ pattern $>$ SAST)
\ENDFOR
\RETURN $F$
\end{algorithmic}
\end{algorithm}

\textbf{Security Structural Regression Detection (\ssr{})}: Algorithm~\ref{alg:ssr} identifies endpoints lacking authentication guards present on sibling routes. Routes are clustered by resource segment, public endpoints are filtered, and majority-rule analysis flags routes where a dominant guard ($\geq$90\% coverage) is absent on destructive HTTP methods (POST, PUT, DELETE, PATCH).

\begin{algorithm}[h]
\caption{Security Structural Regression Detection (\ssr{})}
\small
\label{alg:ssr}
\begin{algorithmic}[1]
\REQUIRE Routing graph $\mathcal{G}$ with routes $\{(path, method, guards)\}$
\ENSURE Set of security regression findings $F$

\STATE \textbf{Filter and cluster}
\FOR{each route $r$ in $\mathcal{G}$}
    \STATE Assign $r$ to cluster $C_{\text{resource}}$ by path segment
\ENDFOR
\STATE Remove clusters matching public whitelist (health, auth, docs, webhook, metrics)

\STATE \textbf{Majority-rule guard analysis}
\STATE $F \leftarrow \emptyset$
\FOR{each remaining cluster $C$ with $|C| \geq 4$}
    \STATE $g^* \leftarrow$ most common guard in $C$
    \STATE $\text{ratio} \leftarrow |\{r \in C : g^* \in \text{guards}(r)\}| / |C|$
    \IF{$\text{ratio} \geq 0.9$}
        \FOR{each route $r \in C$ where $g^* \notin \text{guards}(r)$}
            \IF{method($r$) $\in$ \{POST, PUT, DELETE, PATCH\}}
                \STATE $F \leftarrow F \cup \{(r, g^*, C)\}$
            \ENDIF
        \ENDFOR
    \ENDIF
\ENDFOR
\RETURN $F$
\end{algorithmic}
\end{algorithm}

\sloppy
\smallskip\noindent\textbf{Illustrative Example}: We trace a real-world AI-generated repository through the verification pipeline to illustrate how structural failures manifest and evade standard toolchains.

\begin{tcolorbox}[
  colback=blue!3,
  colframe=blue!60,
  title={\textbf{Illustrative Example: \texttt{hypertropher-app}} \cite{hypertropher-app}},
  fonttitle=\small,
  boxrule=0.5pt,
  arc=2pt,
  left=5pt, right=5pt, top=4pt, bottom=4pt
]

\small
A Next.js/React web application built with AI coding tools and published on GitHub.

\smallskip
\centering
\begin{tabular}{@{}c@{\hspace{12pt}}c@{\hspace{12pt}}c@{\hspace{12pt}}c@{}}
\tiny
\textbf{72} files analyzed &
\tiny
\texttt{tsc-strict} \textcolor{green!60!black}{\checkmark} &
\tiny
SAST \textcolor{green!60!black}{\checkmark} &
\tiny
\textbf{11} structural failures \textcolor{red!70!black}{\boldmath$\times$}
\end{tabular}

\smallskip
\raggedright

\begin{tcolorbox}[
  colback=red!4, colframe=red!40,
  boxrule=0.3pt, arc=1pt,
  left=4pt, right=4pt, top=2pt, bottom=2pt,
  leftrule=3pt
]
\small\textbf{\bci{} --- Configuration Incoherence \hfill 7 findings}\\[2pt]
\scriptsize
Four environment variables (\texttt{NEXT\_PUBLIC\_SUPABASE\_URL}, \texttt{NEXT\_PUBLIC\_SUPABASE\_ANON\_KEY}, \texttt{SUPABASE\_SECRET\_API\_KEY}, \texttt{NEXT\_PUBLIC\_GOOGLE\_MAPS\_API\_KEY}) accessed without defaults across three Supabase client files and the application layout. None declared in any \texttt{.env} or config file. Each resolves to \texttt{undefined} at runtime. Invisible to \texttt{tsc} because \texttt{process.env} access is structurally valid regardless of key existence.
\end{tcolorbox}

\vspace{2pt}

\begin{tcolorbox}[
  colback=violet!4, colframe=violet!40,
  boxrule=0.3pt, arc=1pt,
  left=4pt, right=4pt, top=2pt, bottom=2pt,
  leftrule=3pt
]
\small\textbf{\dhi{} --- Dependency Hallucination \hfill 1 finding}\\[2pt]
\scriptsize
Import references \texttt{@vercel/analytics/next}, a package absent from \texttt{package.json}. The path alias filter correctly excludes 98 \texttt{@/components/ui/*} local aliases, isolating the single genuinely unresolvable external dependency.
\end{tcolorbox}

\vspace{2pt}

\begin{tcolorbox}[
  colback=blue!4, colframe=blue!40,
  boxrule=0.3pt, arc=1pt,
  left=4pt, right=4pt, top=2pt, bottom=2pt,
  leftrule=3pt
]
\small\textbf{\rcf{} --- Resource Coherence \hfill 2 findings}\\[2pt]
\scriptsize
Functions \texttt{loadingCities} and \texttt{previewUrl} declare return types but contain conditional branches that never return a value. CFG reachability analysis identifies the gaps. The type checker misses these because exception flow masks the incomplete returns.
\end{tcolorbox}

\vspace{2pt}

\begin{tcolorbox}[
  colback=teal!4, colframe=teal!40,
  boxrule=0.3pt, arc=1pt,
  left=4pt, right=4pt, top=2pt, bottom=2pt,
  leftrule=3pt
]
\small\textbf{\cfc{} --- Control Flow Coherence \hfill 1 finding}\\[2pt]
\scriptsize
Dead code after a \texttt{return} statement at line 461. This category was absent from all 336 controlled generations yet appears in real-world AI-generated code, consistent with the hypothesis that less supervised generation surfaces failure modes that controlled experiments do not elicit.
\end{tcolorbox}

\smallskip
\small
A companion repository, \texttt{VoiceTradeWithSchwab} \cite{voicetradewithschwab} (voice-controlled stock trading, 100\% AI-generated Python), exhibits 92 findings across five categories including phantom imports, an infinite loop, and unguarded trading configuration variables.

\end{tcolorbox}

\section{Evaluation and Results} \label{sec:evaluation}

\subsection{Experimental Setup}\label{sec:setup}
We evaluate structural failure detection across 336 code generations from two frontier models, GPT-4o (2024-08-06, 128K context) and Claude 3.5 Sonnet (2024-10-22, 200K context), both at temperature zero. The evaluation corpus consists of 10 curated open-source production repositories spanning Python (Django \cite{django}, FastAPI \cite{fastapi}) and TypeScript (Express \cite{expressjs}, Next.js \cite{nextjs}), selected for active maintenance, with a minimum 50 files and 10K LOC, type annotation coverage exceeding 50\%, test coverage above 60\%, and explicit schema definitions.\footnote{Please refer to \cite{patchwork-repo} for the complete list of curated repositories, external validation datasets, and all evaluation metadata.} From these repositories, we extract 60 tasks derived from merged pull requests and closed issues at three complexity levels, namely L1 single-file (30 tasks), L2 multi-file (20 tasks), and L3 cross-cutting (10 tasks). Four prompting strategies control context richness ranging from P1 (minimal, task description only) through P2 (local, 2--5 same-directory files) and P3 (retrieved, 10 similarity-ranked files) to P4 (oracle, 5--15 ground-truth files). The evaluation has a partially unbalanced design in which 24 early tasks were evaluated with GPT-4o under P1 and P2 only, while the remaining 36 tasks received both models across all four strategies, yielding 192 GPT-4o generations (60 each for P1/P2, 36 each for P3/P4) and 144 Claude generations (36 per strategy). P1 and P2 therefore contain 96 generations each and P3 and P4 contain 72 each; all analyses use appropriate denominators to account for this asymmetry.

We compare our framework against four baselines, representing static CI checks: type checking (\texttt{mypy} \cite{mypy} and \texttt{tsc}, test execution, SAST via \texttt{bandit} \cite{bandit} and \texttt{semgrep} \cite{semgrep}, and regex heuristics. Dependency installation is excluded as the evaluation operates on generated code before environment builds; it would catch at most the 3 DHI findings but none of the remaining 64. Detection metrics include per-category precision against ground truth and evasion rates quantifying findings that pass each baseline undetected. Ground truth labels were established through two approaches. For categories with small finding counts (\bci{}, \dhi{}, \pia{}, \srf{}), every finding was manually reviewed and verified as a provable constraint violation. For categories with larger counts (\rcf{}, \ccv{}), precision was established through iterative pipeline refinement that systematically eliminated false positive patterns, with boundary cases resolved by consulting the formal invariants for each category.

\subsection{Results}

\label{sec:results}

\textbf{Detection Performance}: Table \ref{tab:detection_performance} reports detection results across 336 generations (192 GPT-4o, 144 Claude 3.5 Sonnet) under four prompting conditions. Our framework identifies 67 structural failures across eight active categories, with 65 (97.0\%) invisible to all baseline methods. Compilation and type checking detect only 2 findings independently (both \rcf{} return-type violations), while test execution, SAST, and regex heuristics detect none. By category, \rcf{} accounts for the most findings (29), followed by \ccv{} (18), \bci{} (12), \dhi{} (3), \pia{} (3), and \srf{} (2). Manual validation confirms 100\% precision for \bci{}, \dhi{}, \pia{}, and \srf{} (20 of 20 verified as provable constraint violations), while \rcf{} and \ccv{} precision was established through iterative pipeline refinement. No baseline method detects any \ccv{}, \bci{}, \dhi{}, \pia{}, or \srf{} finding; type checkers catch only 2 of 29 \rcf{} findings. Evasion rates reinforce this gap: 97.0\% of findings evade compilation (\texttt{mypy --strict}/\texttt{tsc --strict}), and 100\% evade test suites and SAST tools.

Two categories, \cfc{} and \ssr{}, produced zero findings in the controlled evaluation despite having active detectors. To determine whether these detectors function correctly or whether controlled generation simply does not elicit these failure modes, we applied the full pipeline to 43 real-world AI-generated repositories spanning vibe-coded projects (Cursor AI, Google Gemini, GitHub Copilot), GPT-Engineer/Lovable applications, and self-declared fully AI-generated projects.\footnote{Please refer to \cite{patchwork-repo} for the complete list of all repositories with per-repo metadata and finding counts.} Across 1,581 analyzed files the pipeline detected 1,152 findings in 35 of 43 repositories (81.4\% repo-level incidence), with 474 \dhi{} findings, 270 \rcf{} findings, 177 \pia{} findings, 148 \bci{} findings, 62 \srf{} findings, 16 \cfc{} findings (6 duplicate switch cases, 2 dead-code-after-return, 2 infinite loops), and 5 \ccv{} findings. The \cfc{} findings confirm that the hybrid three-layer detector functions correctly; controlled frontier generation with explicit task specifications simply does not produce the unstructured code patterns that trigger control flow failures. \ssr{} remained at zero across all evaluations. This is consistent with two properties of the evaluation corpus. The detector requires route clusters with at least 4 endpoints exhibiting a dominant per-route guard pattern, and most vibe-coded projects apply authentication at the framework level (e.g., global middleware, app-level decorators) rather than per-route, leaving no inconsistency to detect. The two highest-finding repositories, \texttt{hypertropher-app} and \texttt{VoiceTradeWithSchwab} (detailed in the Illustrative Example above), exemplify how structural failures cluster and compound in real-world AI-generated code.

\begin{table}[h]
\centering
\tiny
\caption{Detection results across 336 generations}
\begin{tabular}{@{}lcccc@{}}
\toprule
\textbf{Method} & \textbf{Findings} & \textbf{TP} & \textbf{Precision} & \textbf{Unique} \\
\midrule
Type Check/Lint & 2 & 2 & 100\% & 0 \\
Test Execution & 0 & 0 & N/A & 0 \\
SAST & 0 & 0 & N/A & 0 \\
Regex Heuristics & 0 & 0 & N/A & 0 \\
\midrule
Patch Work Framework (Ours) & 67 & 67 & see text & 65 \\
\bottomrule
\end{tabular}
\label{tab:detection_performance}
\end{table}

\begin{table}[h]
\centering
\tiny
\caption{Per-category detection results}
\begin{tabular}{@{}lcccl@{}}
\toprule
\textbf{Category} & \textbf{N} & \textbf{Precision} & \textbf{Baseline} & \textbf{Detection Strategy} \\
\midrule
\rcf{} & 29 & Refined & 2/29 & CFG return-path + schema \\
\ccv{} & 18 & Refined & 0/18 & Cross-graph disconnect \\
\bci{} & 12 & 100\% & 0/12 & Unsafe-access + config-space \\
\dhi{} & 3 & 100\% & 0/3 & Registry validation \\
\pia{} & 3 & 100\% & 0/3 & Cross-graph phantom check \\
\srf{} & 2 & 100\% & 0/2 & Import + symbol resolution \\
\cfc{} & 0 & N/A & N/A & Hybrid 3-layer \\
\ssr{} & 0 & N/A & N/A & Resource-clustered auth \\
\bottomrule
\end{tabular}

\label{tab:category_performance}
\end{table}

\begin{figure}[h]
\centering
\includegraphics[width=0.7\linewidth]{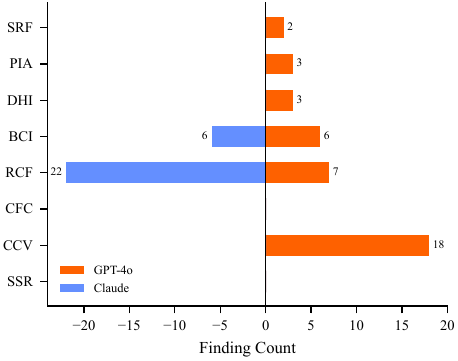}
\caption{Model divergence in structural failure categories. Bars extend left for GPT-4o and right for Claude 3.5 Sonnet.}
\label{fig:model_divergence}
\end{figure}

\textbf{Model Comparison}: Failure distributions diverge qualitatively between GPT-4o and Claude 3.5 Sonnet as visualized in Figure~\ref{fig:model_divergence}. Overall failure rates are comparable (GPT-4o: 39 findings in 25/192 generations, 13.0\%; Claude: 28 findings in 17/144 generations, 11.8\%), but the models exhibit distinct failure profiles rather than simply differing in magnitude. GPT-4o produces all 18 \ccv{} findings and all import-related failures (\dhi{}, \pia{}, \srf{}) exclusively, while Claude generates 22 of 29 \rcf{} findings. Both contribute equally to \bci{} (6 each). Chi-squared testing confirms distributional independence ($\chi^2 = 25.1$, $p = 2.73 \times 10^{-7}$). With 67 total findings, these patterns are descriptive observations warranting replication rather than definitive model characterizations.

\begin{figure}[h]
\centering
\includegraphics[width=0.7\linewidth]{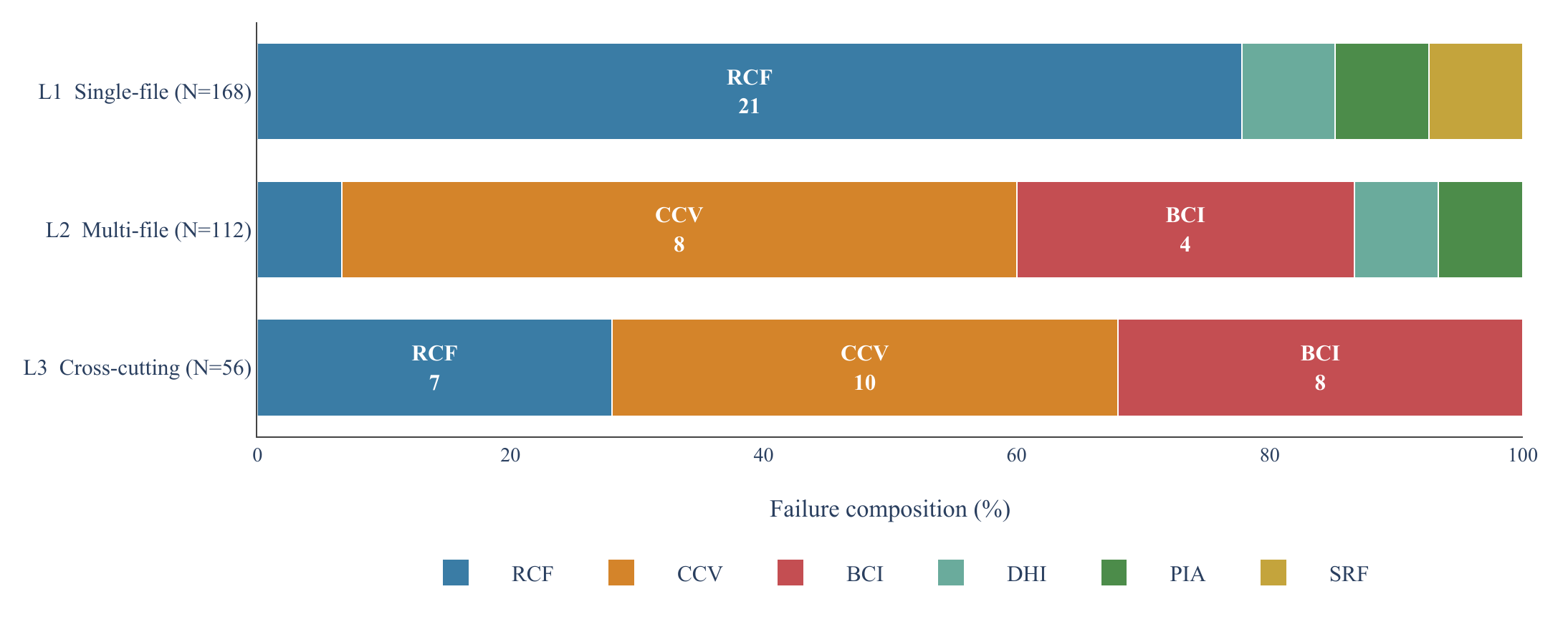}
\caption{Failure category composition by task complexity level.}
\label{fig:complexity_composition}
\end{figure}

\textbf{Prompt Sensitivity}: Table \ref{tab:prompt_sensitivity} reports failure counts by prompting strategy. P3 (retrieved context) exhibits the highest count (24) and P1 (minimal) the lowest (8), indicating that richer context reshapes rather than uniformly reduces failure distributions. Notably, \bci{} appears in P1 through P3 but not P4 (oracle), consistent with ground-truth files helping models identify correct configuration variables. L3 (cross-cutting) tasks exhibit 44.6\% finding incidence compared to 16.1\% for L1 and 13.4\% for L2, confirming that tasks spanning configuration, middleware, and schema boundaries are substantially more failure-prone. Figure~\ref{fig:complexity_composition} shows how failure category composition shifts across complexity levels, with \bci{} and \ccv{} concentrated in L3 tasks that require cross-layer reasoning.

\begin{table}[h]
\centering
\tiny
\caption{Findings by prompt strategy and category}
\begin{tabular}{@{}lcccc@{}}
\toprule
\textbf{Category} & \textbf{P1} & \textbf{P2} & \textbf{P3} & \textbf{P4} \\
\midrule
\rcf{} & 0 & 8 & 13 & 8 \\
\ccv{} & 4 & 5 & 5 & 4 \\
\bci{} & 4 & 2 & 6 & 0 \\
\dhi{} & 0 & 1 & 0 & 2 \\
\pia{} & 0 & 1 & 0 & 2 \\
\srf{} & 0 & 0 & 0 & 2 \\
\midrule
\textbf{Total} & 8 & 17 & 24 & 18 \\
\bottomrule
\end{tabular}

\label{tab:prompt_sensitivity}
\end{table}

\textbf{Runtime}: The pipeline analyzes each file in a median of 47 ms end-to-end (graph construction through all seven detectors), with a 120-second timeout per external analysis script. Registry validation queries (PyPI/npm) are cached across runs. The largest repository in Track D (435 files, 70K LOC) completes in 233 seconds. Runtime is dominated by graph construction (99.8\% of per-file time); all detectors combined complete in under 1ms per file. These times indicate the framework is practical as a CI integration for repositories of moderate size.

\subsection{Practical Implications}

These findings have direct consequences for teams adopting LLM-generated code. First, standard CI pipelines (type checking, testing, SAST) are insufficient as quality gates for generated code; 97\% of detected failures pass all four baselines, meaning structurally broken code can merge undetected. Teams relying solely on existing toolchains face a growing blind spot as LLM-generated code volume increases. Second, the qualitative divergence between models suggests that switching or combining models does not uniformly reduce risk; different models produce different failure profiles, and mitigation strategies should be model-aware. Third, the concentration of failures in L3 cross-cutting tasks (44.6\% incidence versus 13--16\% for simpler tasks) indicates that structural verification is most critical for tasks spanning configuration, routing, and schema boundaries, precisely the tasks where LLMs are increasingly deployed. The framework's localized evidence traces are designed to support both automated CI integration and developer review workflows, providing actionable diagnostics rather than opaque pass/fail signals.

\section{Conclusion}
\label{sec:conclusion}

This work formalized the patchwork problem in LLM-generated code through a graph-based failure taxonomy and a hybrid verification framework combining mature static analysis tools with purpose-built detectors. Across controlled generations and real-world repositories, the overwhelming majority of detected structural failures evade type checking, testing, and SAST entirely, and failure patterns diverge qualitatively between models. Our future work will focus on extending the framework to additional models and programming languages and on subjecting the categories currently validated through iterative refinement to independent precision audits. Further directions include repair mechanisms that leverage detected constraint violations to prompt for missing declarations, and integration into agentic coding workflows and continuous integration pipelines where incremental patch review replaces complete-file generation.


\bibliographystyle{ieeetr}
\bibliography{ref}

\end{document}